\documentclass[twocolumn,floatfix,prx,aps,showpacs]{revtex4-2}
\usepackage{graphicx,amsmath,amssymb,color}
\usepackage{nicefrac}
\usepackage{multirow, makecell, ragged2e}
\usepackage[titletoc,title]{appendix}
\usepackage{float}

\newcommand{\be}{\begin{equation}}
\newcommand{\ee}{\end{equation}}

\newcommand{\ba}{\begin{eqnarray}}
\newcommand{\ea}{\end{eqnarray}}

\renewcommand{\vec}[1]{\mbox{\boldmath$#1$}}

\def\beq{\begin{eqnarray}}
\def\eeq{\end{eqnarray}}

\newcommand\ie{{\it i.e.}~}

\newcommand{\sh}{\mathcal{S}}

\newcommand{\elliptic}[5][\scriptstyle]{\vartheta\left[\begin{array}{c}{{#1 #2}}\\{#1 #3}\end{array}\right]\left(#4\middle|#5\right)}

\makeatletter
\newcommand*{\rom}[1]{\expandafter\@slowromancap\romannumeral #1@}
\makeatother

\begin{document}

\title{Composite anyons on a torus}

\author{Songyang Pu and J. K. Jain}
\affiliation{Department of Physics, 104 Davey Lab, Pennsylvania State University, University Park, Pennsylvania 16802,USA}

\date{\today}
\begin{abstract} 
An adiabatic approach put forward by Greiter and Wilczek interpolates between the integer quantum Hall effects of electrons and composite fermions by varying the statistical flux bound to electrons continuously from zero to an even integer number of flux quanta, such that the intermediate states represent anyons in an external magnetic field with the same ``effective" integer filling factor. 
We consider such anyons on a torus, and construct representative wave functions for their ground as well as excited states. These wave functions involve higher Landau levels in general, but can be explicitly projected into the lowest Landau level for many parameters. We calculate the variational energy gap between the first excited state and ground state and find that it remains open as the statistical phase is varied. Finally, we obtain from these wave functions, both analytically and numerically, various topological quantities, such as ground-state degeneracy, the Chern number, and the Hall viscosity. 

\end{abstract}
\maketitle

\section{introduction}
Composite fermions \cite{Jain89}, the topological bound states of electrons and an even number ($2s$) of quantized vortices, lead to an explanation of the fractional quantum Hall effect \cite{Tsui82,Prange87} at fractions $\nu=n/(2sn\pm 1)$ as the integer quantum Hall effect of composite fermions, and they allow a calculation of the topological and non-topological features of these fractional quantum Hall states\cite{DasSarma07,Heinonen98,Jain07,Halperin20}. 
Greiter and Wilczek \cite{Greiter90,Greiter92b,Greiter21} proposed an adiabatic approach, wherein the fractional quantum Hall effect is connected to the integer quantum Hall effect by continuously tuning the strength of the vortex attached to electrons from zero to $2s$, while at the same time varying the external magnetic field in such a manner that the effective magnetic field remains unchanged. The two limiting cases are familiar and well-studied. When the number of attached vortices is zero, we of course have the integer quantum Hall effect of non-interacting electrons. When the number of attached vortices is an even integer, the base particles are composite fermions (CFs), producing the fractional quantum Hall effect of electrons at $\nu=n/(2sn\pm 1)$. This article is concerned with the intermediate states, when the number of vortices attached to each particle is a rational fraction and the base particles are anyons obeying fractional braiding statistics \cite{Leinaas77,Wilczek82}. Recently, a Chern-Simons field theory with fluctuating dynamical gauge field has also been used to forge a bridge connecting the integer and fractional quantum Hall states~\cite{Hansson21}.

We note that the excitations of the fractional quantum Hall states have been predicted to obey fractional braiding statistics~\cite{Halperin84,Arovas84,Jain07}.
In contrast, we are dealing in our study with fictitious anyons designed to interpolate between the integer and fractional quantum Hall states.

The theoretical studies on anyons have been attempted through various methods, including field theories \cite{Iengo92,Chen89,Iengo91,Iengo90,Hansson96}, exact diagonalization \cite{Sporre91,Sporre93,Canright89,Canright89a,Hanna89,Xie90,Hatsugai91,Kudo20,Ouvry19}, density functional theory \cite{Hu21}, wave functions \cite{Wu84,Thouless85,Girvin90,Chin92,Lundholm17,Lundholm13,Fayyazuddin93,Fayyazuddin93b}, and other methods \cite{Laughlin88,Fetter89,Wen90c,Lee91,Chitra92,Li92,Correggi17}. The wave function approach gives an explicit description of the many-particle states and allows direct calculations of both topological and non-topological physical quantities. Earlier, the wave function studies were mainly based on disk geometry \cite{Wu84,Girvin90,Chin92,Lundholm17,Lundholm13}.

In this work, we revisit the problem of constructing an anyon wave function on a torus for general filling factors. The torus geometry offers certain special advantages. One of them is that the torus is compact, which avoids complications from edge states hosted by open boundaries. The shape of the torus and the boundary conditions are tunable, which makes it an ideal geometry to study topological bulk quantities such as Chern number and Hall viscosity. The wave functions for CFs carrying $2s$ vortices have been constructed on a torus in Ref.~\cite{Pu17}. However, unlike in the disk geometry, a generalization from CFs to anyons cannot be accomplished by simply replacing the integer number of attached vortices by a fractional number. The interplay of the periodic boundary conditions and fractional statistics imposes a nontrivial braiding group for anyons on a torus \cite{Einarsson90}, which requires the wave functions for anyons to be multi-component. An alternative way to understand the origin of the multi-component structure has been discussed by Fayyazuddin \cite{Fayyazuddin93}, as arising from the coupling of the gauge field and the particle degrees of freedom. This multi-component structure is consistent with the exact diagonalization results on a lattice Hamiltonian on a torus \cite{Wen92c,Hatsugai91,Kudo20} and also with the Chern-Simons theory \cite{Iengo91,Iengo92,Hosotani92}. Ref.~\cite{Fayyazuddin93b} has studied anyon ground-state wave functions on a torus using the Chern-Simons gauge transformation.

We achieve a construction of trial wave functions for degenerate ground states as well as excited states of anyons for general statistical parameters and filling factors such that the effective filling factor is an integer. These wave functions are more general and have a simpler form than those constructed previously in the literature, reducing to the Jain CF wave functions \cite{Pu17} when the number of attached fluxes to each particle is an even integer. We believe, from experience with the CF theory, that the  lowest Landau level (LLL) projections of these wave functions should provide a good account of anyons interacting by a repulsive interaction, such as the Coulomb interaction. However, we have not investigated the quantitative validity of these wave functions. We calculate below variational excitation gaps, as well as several topological properties of the incompressible states of anyons, which are expected to be insensitive to the details of the wave function.

The remainder of the paper is organized as follows. In Sec.~\ref{wf}, we briefly review the braiding group of anyons on a torus. Then we construct a complete set of multi-component anyon wave functions that provide a representation of this braiding group. We also show that our wave functions have the expected ground state degeneracies. In Sec.~\ref{energy}, we calculate the variational values for the charge gaps for anyons interacting via the Coulomb interaction, and we find that the gap is preserved as we tune the number of attached vortices; this supports the view that the process is adiabatic and is also consistent with exact diagonalization findings by Kudo and Hatsugai\cite{Kudo20}. In Secs.~\ref{Chern number} and ~\ref{Hall viscosities}, we calculate the Chern number and the Hall viscosity analytically and numerically. We find that the total Hall viscosity can be viewed as the sum of the Hall viscosities of different factors in the wave function, and it encodes information on the number of filled effective Landau levels and the anyon statistics. We summarize our results in Sec.~\ref{summary}. Our results are consistent with the work by Kudo and Hatsugai \cite{Kudo20}, who have diagonalized a lattice model Hamiltonian in the torus geometry and numerically calculated the ground-state degeneracies, gaps, and Chern numbers.

\section{multi-component anyon wave functions and ground state degeneracies}
\label{wf}

We consider a two-dimensional many-particle system on the surface of a torus with a perpendicular external magnetic field applied. We assume that there are $N$ anyons and $N_\phi$ external magnetic flux quanta. The filling factor of anyons is $\nu=N/N_\phi$. The adiabatic transport of an anyon around another along a closed loop results in a statistical phase $2\theta$ (which defines our anyon). 
In what follows, we define  
\be
\theta=\pi\left(1+{p\over q}\right)=\pi{p'\over q}
\ee
with $p'=p+q$. 
When ${\theta\over \pi}$ is an odd (even) integer, the particles are fermions (bosons). 
The anyons can be mapped into fermions or bosons 
with gauge fluxes attached. Assuming that the base particles are fermions, 
the number of effective magnetic flux quanta felt by them (counting both the external magnetic field and the statistical field) is 
\be
N_\phi^{f}=N_\phi-\left({\theta\over \pi}-1\right)N=N_\phi-{p\over q}N,
\ee
and the effective filling factor for fermions, $\nu_f$, is given by
\be
{1\over \nu_f}={1\over \nu}-{\theta\over \pi}+1={1\over \nu}-{p\over q}.
\ee 
It reduces to the standard CF theory when $p/q$ is an even integer.
If the base particles are chosen to be bosons instead, the number of effective magnetic flux quanta felt by them is 
\be N_\phi^{b}=N_\phi-{\theta\over \pi}N=N_\phi-{p'\over q}N,
\ee 
and the inverse of the bosonic filling factor is 
\be
{1\over \nu_b}={1\over \nu}-{\theta\over \pi}={1\over \nu}-{p'\over q}.
\ee
(We note that we use a different convention for the definition of $\theta$ compared with that in Ref.~\cite{Kudo20}. Our $\theta$ corresponds to $2\pi-\theta$ of that paper. We choose our convention because it is more natural for CFs, as $\theta$ is simply $m\pi$ at filling $1/m$ for CFs. Our convention is 
consistent with that of Refs.~\cite{Greiter90,Greiter92b}.) 

A torus can be mapped into a parallelogram on a complex plane with quasi-periodic boundary conditions imposed. We define the real axis along one edge of the parallelogram and name the length of that edge as $L_1$. The other edge of the parallelogram is defined as $L_2=L_1\tau$, where $\tau=\tau_1+i\tau_2$ is a complex number called the modular parameter \cite{Gunning62} of the torus. In this work, we assume the external magnetic field is $\vec{B}=-B\hat{z}$. Then, in the symmetric gauge, 
the LLL wave function is a holomorphic function of the particle coordinates $z_i=x_i+iy_i$ times a Gaussian factor $e^{-{|z|^2\over 4\ell^2}}$, where $\ell=\sqrt{\hbar c/eB}$ is the magnetic length. Later we will also use the reduced particle coordinates $\theta_{1,i}, \theta_{2,i}\in [0,1)$ which are defined through $z_i=L_1\theta_{1,i}+L_2\theta_{2,i}$. The total area of the torus is $V=L_1^2\tau_2=2\pi N_\phi \ell^2$. The quasi-periodic boundary conditions under external magnetic field are defined through magnetic translation operators \cite{Zak64,Brown64}. For Hall viscosities, it is convenient to choose the $\tau$ gauge $(A_x,A_y)=B(y,-{\tau_1\over \tau_2}y)$ \cite{Fremling14,Pu20}. In this gauge, the magnetic translation operators acting on a single particle $z=L_1\theta_1+L_2\theta_2$ are defined as:
\be
\label{magnetic translation operator}
t\left(\alpha L_1+\beta L_2\right)=e^{\alpha\partial_1+\beta\partial_2+i2\pi\beta N_{\phi}\theta_1},
\ee
where $\partial_1\equiv{\partial \over {\partial \theta_1}}$ and $\partial_2\equiv{\partial \over {\partial \theta_2}}$ . 

Fermions or bosons satisfy the quasi-periodic boundary conditions on a torus:
\be
\label{pbc}
t(L_i)\psi(z,\bar{z})=e^{i\phi_i}\psi(z,\bar{z}) \quad i=1,2.
\ee
However, this is not the case for anyons since this equation is inconsistent with the fractional statistics. As shown in Ref.~\cite{Birman69}, the braiding can be accomplished on a torus by wrapping two particles along the two periodic loops. If the periodic boundary conditions are just represented by phases, the braiding statistics can only be integer multiples of $\pi$.  Ref.~\cite{Einarsson90,Hatsugai91} have shown that the boundary conditions for anyons on a torus are given by:
\be
\label{bg1}
t_j(L_1)\Psi=e^{i\phi_1}e^{-i2j\theta}
\begin{pmatrix}
1&\cdots&\\
&c&\cdots\\
\vdots&\vdots&\vdots\\
\cdots&&c^{q-1}\\
\end{pmatrix}
\Psi
\ee
\be
\label{bg2}
t_j(L_2)\Psi=e^{i\phi_2}e^{i2j\theta}
\begin{pmatrix}
0&1&0&\cdots&0\\
0&0&1&\cdots&0\\
\vdots&\vdots&\vdots&\vdots\\
1&0&0&\cdots&0\\
\end{pmatrix}
\Psi,
\ee
with $\theta=\pi{p'\over q}$, $c=e^{i2\pi{p'\over q}}$. The twisted boundary conditions are defined by phases $\phi_1$ and $\phi_2$, which represent $2\pi$ times the number of magnetic flux quanta through the two holes of the torus \cite{Hatsugai91}. 
Here $t_j$ is the magnetic translation operator acting on the $j$th particle. $\Psi$ is a $q$-component vector. 
For the special case of CFs, $q=1$, and the above boundary conditions are satisfied by the Jain CF wave functions constructed in Ref.~\cite{Pu17}.

In the remaining part of this section, we first find a solution for Eq.~\ref{bg1} and Eq.~\ref{bg2} for ${1\over \nu_f}=1$ by ansatz, just as was done in Ref.~\cite{Haldane85} for Laughlin wave functions. Then we show how to generalize the solution to other fillings. Finally, we show how to obtain wave functions for all degenerate ground states.

\subsection{$\nu_f=1$}

The effective filling factor $\nu_f=1$ is obtained when anyons with statistics $\theta=\pi{p'\over q}$ have a filling factor $\nu={q\over p'}$.  
Here, the fermions fill the lowest Landau level, and the ground state 
 has $q$ components. In what follows, we obtain trial wave functions for the ground and excited states.

We make the following ansatz for the wave function (we use superscript $\alpha$ to label the degeneracy and subscript $k$ to label the component):
\be
\label{anyon wf}
\Psi^{(\alpha)}=(\Psi^{(\alpha)}_0,\Psi^{(\alpha)}_1,\Psi^{(\alpha)}_2,\cdots,\Psi^{(\alpha)}_{q-1})^T
\ee
\be
\label{wf2}
\Psi^{(\alpha)}_k[z_i]=e^{i\pi\tau N_\phi\sum_i \theta_{2,i}^2}\prod_{i=1}^NJ_iF^{(\alpha)}_k(Z),
\ee
\be
\label{J1}
J_i=\prod_{j>i}e^{{p'\over q} \ln \elliptic{1/2}{1/2}{{z_i-z_j\over L_1}}{\tau}}.
\ee 
Here $Z=\sum_{i=1}^Nz_i$ is the center-of-mass coordinate, and we use the Jacobi theta function with rational characteristics\cite{Mumford07}; its definition and some of its properties are listed in Appendix \ref{theta-function}.  The ansatz wave function has three parts. The first part $e^{i\pi\tau N_\phi\sum_i \theta_{2,i}^2}$ 
appears in the $\tau$ gauge. The second part $\prod_{i=1}^NJ_i$ is purely made up of the relative coordinates $z_i-z_j$; it is analogous to the Jastrow factor $\prod_{i<j}\left(z_i-z_j\right)$ in the disk geometry. The 
coefficient ${p'\over q}$ in the exponential of Eq.~\ref{J1} is fixed by the braiding statistics and defines the number of attached vortices (it is replaced by an even integer for CFs). The last part $F^{(\alpha)}_k(Z)$ is the center-of-mass part, which carries both the degeneracy index and component index. In making this ansatz, we assume that the wave function can be written as a product of the relative part and the center-or-mass part, 
which is known to be true for $q=1$, \ie, for the Laughlin states.

Note that Eq.~\ref{J1} has a branch cut. We adopt the convention that when $z\rightarrow z\pm L_1$, $\ln \elliptic{1/2}{1/2}{{z\over L_1}}{\tau}\rightarrow \ln \elliptic{1/2}{1/2}{{z\over L_1}}{\tau}\pm i\pi$, \ie, when the particle coordinate moves to the right (left) across the boundary, the $\ln \vartheta$ function goes up (down) on the Riemann surface.

Now we need to solve for the center-of-mass part $F^{(\alpha)}_k(Z)$. According to Eq.~\ref{bg1} and Eq.~\ref{bg2}, it satisfies
\be 
\label{Fbg1}
F^{(\alpha)}_k(Z+L_1)=e^{i\left[\phi_1+2\pi{p'\over q}\left(k-{N+1\over 2}\right)\right]}F^{(\alpha)}_k(Z)
\ee
\be
\label{Fbg2}
F^{(\alpha)}_k(Z+L_2)=e^{-i\left({2\pi p'Z\over qL_1}-\phi_2-{p'\over q}\pi(N+1)+{\pi p' \tau\over q}\right)}F^{(\alpha)}_{k+1}(Z).
\ee
with $k=0,1,\cdots q-1$ and $F^{(\alpha)}_q(Z)=F^{(\alpha)}_0(Z)$.
We can do a Fourier expansion of $F^{(\alpha)}_k(Z)$ according to Eq.~\ref{Fbg1}
\be
F^{(\alpha)}_k(Z)=\sum_n d_{k,n}^{(\alpha)}e^{i\left(2\pi n+\phi_1+2\pi{p'\over q}\left(k-{N+1\over 2}\right)\right){Z\over L_1}}.
\ee
Through Eq.~\ref{Fbg2} and $F^{(\alpha)}_q(Z)=F^{(\alpha)}_0(Z)$, the coefficients are fixed as
\be
d_{k+1,n}^{(\alpha)}=e^{i\left(2\pi n+\phi_1+2\pi{p'\over q}\left(k-{N\over 2}\right)\right)\tau-i\left(\phi_2+{p'\over q}\pi(N+1)\right)}d_{k,n}^{(\alpha)}
\ee
\be
d_{0,n+p'}^{(\alpha)}=e^{i\left(2\pi nq+\pi p'q+q\phi_1-p'\pi(N+1)\right)\tau-i(q\phi_2+p'\pi (N+1))}d_{0,n}^{(\alpha)}
\ee
Because there are $p'$ independent coefficients, evidently, this tells us that there are $p'$ independent solutions, depending on our choice of the coefficients $d_{0,0}^{(\alpha)}$, $d_{0,1}^{(\alpha)}$, $\cdots$ $d_{0,p'-1}^{(\alpha)}$. These solutions can be written in an elegant form using theta functions:
\be
\label{FK}
F_k^{(\alpha)}(Z)=\elliptic{a_k}{b_\alpha}{Z\over L_1}{{q\over p'}\tau}
\ee
\be
a_k={1\over 2\pi}\left(\phi_1-{p'\over q}\pi (N+1)+2\pi k{p'\over q}\right)
\ee
\be
b_\alpha=-{1\over 2\pi}\left({q\over p'}(\phi_2+2\pi\alpha)-2\pi(N-1){q\over p'}+\pi(N+1)\right)
\ee
with $k=0,1,\cdots q-1$ and $\alpha=0,1,\cdots p'-1$. Eq.~\ref{anyon wf}, Eq.~\ref{wf2}, Eq.~\ref{J1}, and Eq.~\ref{FK} together give the $p'$-fold degenerate $q$-component ground state wave functions.

If we apply the center-of-mass magnetic translation $t_{\rm CM}\left(L_2/N_\phi\right)=\prod_{i=1}^Nt_i\left(L_2/N_\phi\right)$ on $\Psi^{(\alpha)}$, we get
\ba
t_{\rm CM}\left(L_2/N_\phi\right)\Psi^{(\alpha)}&=&e^{i{q\over p'}\left(\phi_2+2\pi(\alpha-N+1)+{p'\over q}\pi (N+1)\right)}\Psi^{(\alpha)} \nonumber\\
\ea
The degenerate states have different eigenvalues, and hence are orthogonal. On the other hand, they can be transformed into one another by applying $t_{\rm CM}\left(L_1/N_\phi\right)$:
\ba
t_{\rm CM}\left(L_1/N_\phi\right)\Psi^{(\alpha)}&=&\Psi^{(\alpha-1)}
\ea

One may notice that for the special case of $\nu=1/m$, which corresponds to the Laughlin state for fermions or bosons, Eq.~\ref{anyon wf} does not have the familiar form given in general literature. (For instance, one can compare Eq.~\ref{anyon wf} to Eq. 6 in Ref.~\cite{Pu20b}.) Actually, they are related by an $m$-dimensional unitary transformation. While Eq.~\ref{anyon wf} is an eigenstate of $t_{\rm CM}\left(L_2/N_\phi\right)$, the more familiar Laughlin wave function (e.g., see Ref.~\cite{Pu20b}) is chosen to be the eigenstates of $t_{\rm CM}\left(L_1/N_\phi\right)$. If one defines the periodic properties Eq.~\ref{bg1} and Eq.~\ref{bg2} such that $t_n(L_2)$ is diagonal and $t_n(L_1)$ is non-diagonal, the more familiar form will be recovered.

\subsection{$\nu_f=n$}

Now let us consider the case of more general $\nu_f=n$. This corresponds to the case of $N$ anyons with statistical parameter $\theta=\pi(1+{p\over q})$, in a magnetic field with flux number $N_\phi=({1\over n}+{p\over q})N$, at $\nu={nq \over q+np}$. 
 When we model the anyons as fermions with ${p\over q}$ vortices attached to them, the fermions fill $n$ Landau levels in the effective magnetic field, \ie $\nu_f=n$. 
In this case, we construct below the wave function as a product of a $q$-component anyon wave function and the fermionic or bosonic scalar wave function in the effective magnetic field. We show that the ground state degeneracy is given by $q+np$. (For $n=1$, which corresponds to $\nu_f=1$, this gives a ground state degeneracy of $q+p=p'$, consistent with the previous subsection.)

Following the standard CF construction, we first write the wave function as a product state:
\begin{widetext}
\be
\label{anyon wf2}
\Psi_{n;{p\over q}}=\Psi_n\Psi_{p\over q}^{(N-1)}
\ee
Here, the $q$-component $\Psi_{p\over q}^{(N-1)}$ is given by Eq.~\ref{anyon wf} with $N_\phi\rightarrow {p\over q}N$, $p'\rightarrow p$, $\phi_1\rightarrow \phi_1^a$, $\phi_2 \rightarrow \phi_2^a$. We choose $\alpha=N-1$ just to simplify the phase factor under $t_{\rm CM}\left({qL_2\over pN}\right)$. The  other part $\Psi_n$ is the (single-component) wave function of $n$-filled Landau levels \cite{Pu20}:
\be
\label{eq:Psi_n}
\Psi_n[z_i,\bar{z}_i]=e^{i\pi\tau N_\phi^* \sum_{j=1}^{N}\theta_{2,j}^2}{1\over \sqrt{N}}\chi_n[f_{i}(z_j,\bar{z}_j)].
\ee
\be
\label{upro chi}
\chi_m[f_{i}(z_j,\bar{z}_j)]=
\begin{vmatrix}
f_0^{(0)}(z_1)&f_0^{(0)}(z_2)&\ldots&f_0^{(0)}(z_N) \\
f_0^{(1)}(z_1)&f_0^{(1)}(z_2)&\ldots &f_0^{(1)}(z_N)\\
\vdots&\vdots&\vdots \\
f_0^{(N_\phi^*-1)}(z_1)&f_0^{(N_\phi^*-1)}(z_2)&\ldots&f_0^{(N_\phi^*-1)}(z_N) \\
f_1^{(0)}(z_1,\bar{z}_1)&f_1^{(0)}(z_2,\bar{z}_2)&\ldots&f_1^{(0)}(z_N,\bar{z}_N) \\
f_1^{(1)}(z_1,\bar{z}_1)&f_1^{(1)}(z_2,\bar{z}_2)&\ldots &f_1^{(1)}(z_N,\bar{z}_N)\\
\vdots&\vdots&\vdots \\
f_{n-1}^{(N_\phi^*-1)}(z_1,\bar{z}_1)&f_{n-1}^{(N_\phi^*-1)}(z_2,\bar{z}_2)&\ldots&f_{n-1}^{(N_\phi^*-1)}(z_N,\bar{z}_N)\\
\end{vmatrix}.
\end{equation}
\be
\label{eq:psi_n_basis}
f_n^{(k)}(z,\bar{z})=
\sum_{t\in\mathbb{Z}+\frac{k}{N_{\phi}^*}+{\phi_1^f\over 2\pi N_\phi^*}}e^{i\pi N_{\phi}^*\tau t^2}e^{i2\pi N_{\phi}^*t\left(\frac{z}{L_1}-{\phi_2^f\over 2\pi N_\phi}\right)}
H_{n}\left(\frac{\tau_{2}L_1}{\ell_{B}}\left(\theta_2+t\right)\right),
\ee
where $N_\phi^*=N/n$ and $H_{n}(x)$ are the Hermite polynomials. (We omit the normalization factors here.)
\end{widetext}
Since the phases generated by magnetic translation operators simply add, the above wave function satisfies Eq.~\ref{bg1} and Eq.~\ref{bg2} on the condition that $\phi_1=\phi_1^f+\phi_1^a$ and $\phi_2=\phi_2^f+\phi_2^a$. 
We find it natural to make the following choice for the phases:
\be
\phi_i^a={np\over np+q}\phi_i, \quad i=1,2
\ee
\be
\phi_i^f={q\over np+q}\phi_i, \quad i=1,2.
\ee
In this choice, the magnetic fields through the two holes of the torus felt by the fermions and anyons are in the same proportion as the magnetic fields perpendicular to the torus felt by the fermions and anyons. We conjecture that for this choice, there exists at least one momentum sector in which the wave function is well defined for all $\phi_1$ and $\phi_2$. We prove this conjecture in Appendix~\ref{appen a} for $\nu_f=1$, and we have found it to be valid for all cases below.

One might at first think that the wave function in Eq.~\ref{anyon wf2} has a $p$-fold degeneracy, in contrast to the expected ($q+np$)-fold degeneracy~\cite{Kudo20}. Below we show how to reproduce the $q+np$ degenerate wave functions from Eq.~\ref{anyon wf2}. First, we note that $\Psi_n$ satisfies:
\be
t_{\rm CM}\left({nL_2\over N}\right)\Psi_n=e^{i(\phi_2^f+\pi(N-1))n}\Psi_n
\ee
while $\Psi_{p\over q}$ satisfies:
\be
\label{pq}
t_{\rm CM}\left({qL_2\over pN}\right)\Psi_{p\over q}^{(N-1)}=e^{i\left({q\over p}\phi_2^a+\pi(N+1)\right)}\Psi_{p\over q}^{(N-1)}.
\ee
Therefore, Eq.~\ref{anyon wf2} satisfies:
\be
t_{\rm CM}\left({qnL_2\over N}\right)\Psi_{n;{p\over q}}=e^{i\left(qn\phi_2+n\pi(N+1)(p+q)\right)}\Psi_{n;{p\over q}}.
\ee
We can define a momentum projection operator:
\be
\label{P}
P_\alpha={1\over \sqrt{np+q}}\sum_{j=0}^{np+q-1}\left[e^{-i{qn\phi_2+n\pi(N+1)(p+q)+2\pi\alpha\over np+q}}t_{\rm CM}\left({L_2\over N_\phi}\right)\right]^j.
\ee
This generates the degenerate eigenstates:
\be
\label{degeneracy}
t_{\rm CM}\left(L_2/N_\phi\right)P_\alpha \Psi_{n;{p\over q}}=e^{i{qn\phi_2+n\pi(N+1)(p+q)+2\pi\alpha\over np+q}}P_\alpha \Psi_{n;{p\over q}}
\ee
where $\alpha=0,1,2,\cdots np+q-1$ corresponds to the $np+q$-fold ground state degeneracy. The $np+q$-fold degenerate states are related to each other by $t_{\rm  CM}\left(L_1/N_\phi\right)$. With some possible gauge transformation, the degenerate states have such relation
\be
t_{\rm CM}\left(L_1/N_\phi\right)P_\alpha \Psi_{n;{p\over q}}=e^{i{nq\phi_1+\pi q(N-n)\over np+q}}P_{\alpha-nq} \Psi_{n;{p\over q}}
\ee
To ensure the wave function after momentum projection does not vanish, we have to choose $\phi_i^f={q\over q+pn}\phi_i$ and $\phi_i^a={np\over q+np}\phi_i$ for $i=1,2$, as explained in Appendix~\ref{appen a}.  Note that the degeneracy is equal to $q+np$, which is equal to the denominator of the filling factor $\nu={nq\over q+np}$ only when $nq$ and $q+np$ are mutually coprime. For example, for $n=2$ and $p/q=1/4$, we have $\nu=4/3$ while the degeneracy is $6$. This result is consistent with the exact diagonalization results shown in Ref.~\cite{Kudo20}.

For $n\ge 2$, $\Psi_{n;{p\over q}}$ given by Eq.~\ref{anyon wf2} is not fully in the LLL. In general, one can apply a direct LLL projection following Ref.~\cite{Girvin84b}. However, the direct-projected wave functions cannot be used to calculate systems typically with more than ten particles. An alternative Jain-Kamilla projection can be applied to evaluate large systems \cite{Jain97,Jain97b}. If ${p\over q}\ge 2$, the modified Jain-Kamilla projection \cite{Pu17,Pu20} can be implemented as:
\be
\label{anyon wf3}
\Psi_{n;{p\over q}}=(\Psi_0,\Psi_1,\Psi_2,\cdots,\Psi_{q-1})^T
\ee
 \be
 \label{projected wf}
 \Psi_k=
 e^{\i\pi\tau N_\phi\sum_i \theta_{2,i}^2}F_k(Z)\prod_{i=1}^N\bar{J}_i
 \chi_n[\hat{g}_{i}(z_j)\tilde{J}_j],
\ee
\be
\label{chi-det}
{\chi_n}[\hat{g}_{i}(z_j)\tilde{J}_j]=
\begin{vmatrix}
\hat{g}_0^{(0)}(z_1)\tilde{J}_1&\ldots&\hat{g}_0^{(0)}(z_N)\tilde{J}_N \\
\vdots&\vdots&\vdots \\
\hat{g}_1^{(0)}(z_1)\tilde{J}_1^p&\ldots&\hat{g}_1^{(0)}(z_N)\tilde{J}_N \\
\vdots&\vdots&\vdots \\
\end{vmatrix},
\ee
\be
\bar{J}_i=\prod_{j\neq i}e^{\left({p\over 2q}-1\right) \ln \elliptic{1/2}{1/2}{{z_i-z_j\over L_1}}{\tau}}.
\ee 
\be
\tilde{J}_i=\prod_{j\neq i}e^{\ln \elliptic{1/2}{1/2}{{z_i-z_j\over L_1}}{\tau}}.
\ee 
where $F_k(Z)$ is given by Eq.~\ref{FK} with $p'\rightarrow p$, $\phi_1\rightarrow \phi_1^a$, $\phi_2 \rightarrow \phi_2^a$. The general form of $\hat{g}_n^{(k)}(z)$ was derived in detail in Refs.~\cite{Pu17,Pu20}. Here we give the form for the lowest three Landau levels (without including any normalization factors):
\begin{widetext}
\be
\hat{g}_0^{(k)}(z)=f_0^{(k)}(z)=\elliptic{{k\over N_\phi^*}+{\phi_1^f\over 2\pi N_\phi^*}}{-{\phi_2^f\over 2\pi}}{N_\phi^* z\over L_1}{N_\phi^* \tau},
\ee
\be
 \label{2nd LL matrix element}
 \hat{g}_1^{(k)}(z)=(N_\phi^*-N_\phi)\frac{\partial f_0^{(k)}(z)}{\partial z}+N_\phi^*f_0^{(k)}(z)2\frac{\partial}{\partial z},
 \ee
\be
\hat{g}_2^{(k)}(z)= (N_\phi-N_\phi^*)^2{\partial^2 f_0^{(k)}(z)\over\partial z^2}
-2N_\phi^*(N_\phi-N_\phi^*){\partial f_0^{(k)}(z)\over\partial z}2{\partial\over \partial z}
+N_\phi^{*2}f_0^{(k)}\left(2{\partial\over \partial z}\right)^2,
\ee
\end{widetext}

We mention a caveat for the projected wave function Eq.~\ref{projected wf}. Compared to the Jastrow factor in the unprojected wave function Eq.~\ref{J1}, we changed $\prod_{i<j}{\rm exp}\left({{p\over q} \ln \elliptic{1/2}{1/2}{{z_i-z_j\over L_1}}{\tau}}\right)$ to $\prod_{i\neq j}{\rm exp}\left({{p\over 2q} \ln \elliptic{1/2}{1/2}{{z_i-z_j\over L_1}}{\tau}}\right)$. For composite fermions, i.e. when $p\over q$ is an even integer, this process only generates a factor of $(-1)^{pN(N-1)\over 4q}$, which is of no significance. However, the Jastrow factors of anyons have branch cuts, and the definition of how the Jastrow factors change across the branch cuts is very subtle. If we use the  definition for the multi-valued Jastrow factors mentioned right after Eq.~\ref{branch}, we find the projected wave function no longer satisfies Eq.~\ref{bg1} and Eq.~\ref{bg2}. On the other hand, if we confine the particles to the principal region of the torus (\ie the parallelogram spanned by $L_1$ and $L_2$), the projected wave function captures the lowest Landau level part of the unprojected wave function, which satisfies the imposed braiding group. Therefore, the projected wave function is still sufficient for calculating local physical quantities such as energies, Berry curvatures, and Hall viscosities.

An important property that is required for wave functions on a torus is modular covariance. We discuss this issue in Appendix~\ref{appen b} and show that the anyon wave function constructed above is modular covariant.

\section{energy gap and the adiabatic principle}
\label{energy}
The key point of the adiabatic principle is that the ground states remain gapped as we tune the strength of the attached vortex, or in other words the statistical phase $\theta$, in such a manner that the effective filling factor remains constant. In this section, we numerically confirm this statement by calculating the transport gaps for $\nu_f=1,2$ using our ansatz wave functions, assuming Coulomb interaction between the anyons.

We calculate the transport gaps by creating a quasiparticle state and a quasihole state separately. A quasiparticle can be obtained from Eq.~\ref{anyon wf3} by occupying an extra orbital in the lowest unoccupied effective Landau level in the Slater determinant part. Similarly, a quasihole can be obtained by leaving an unoccupied orbital in the highest occupied effective Landau level in the Slater determinant part. For $\nu_f=1$, the transport gap is calculated as:
\ba
\label{delta}
\Delta_{n=1;{p\over q}}(N)&=&E^{\rm qp}\left(N,N_\phi=\left(1+{p\over q}\right)N+1\right)\nonumber \\
&&+E^{\rm qh}\left(N,N_\phi=\left(1+{p\over q}\right)N-1\right)\nonumber \\
&&-2E^{\rm 0}\left(N,N_\phi=\left(1+{p\over q}\right)N\right).
\ea
where $E^{\rm qp}$,  $E^{\rm qh}$ and $E^{\rm 0}$ are the energies of the quasiparticle, the quasihole and the ground states. For $\nu_f=2$, the transport gap is calculated as:
\ba
\label{delta2}
\Delta_{n=2;{p\over q}}(N)&=&E^{\rm qp}\left(N-1,N_\phi=\left({1\over 2}+{p\over q}\right)N-{p\over q}\right)\nonumber \\ 
&&+E^{\rm qh}\left(N+1,N_\phi=\left({1\over 2}+{p\over q}\right)N+{p\over q}\right) \nonumber \\ 
&&-2E^{\rm 0}\left(N,N_\phi=\left({1\over 2}+{p\over q}\right)N\right)
\ea
We assume Coulomb interaction between particles.

\begin{figure}[t]
	\includegraphics[width=\columnwidth]{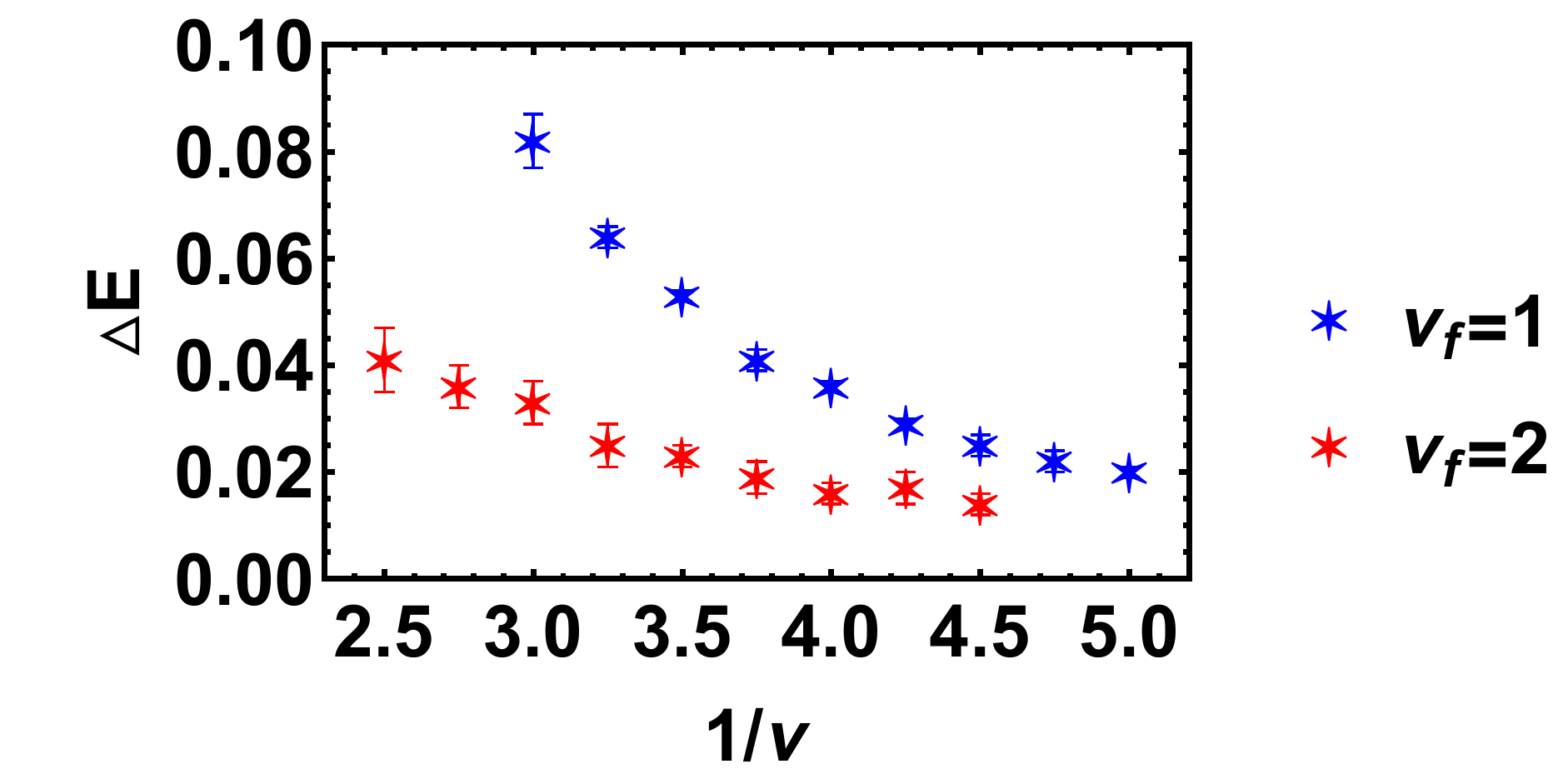} 
	\caption{
The transport gaps for $\nu_f=1$ and $\nu_f$=2 as a function of $1/\nu$ for $N=20$, obtained from lowest-Landau-level projected variational wave functions. Coulomb interaction is assumed between the anyons, and the energies are quoted in units of $e^2/\epsilon \ell$.
 }
	\label{gap}
\end{figure}

We use variational Monte Carlo and the anyon wave function Eq.~\ref{anyon wf3} to calculate the transport gaps. The results are shown in Fig.~\ref{gap}. Because the energy only depends on the relative part of the wave function, we only use the first component in Eq.~\ref{anyon wf3} to calculate the energy and multiply the values by the number of components to save the computation time. For $\nu_f=1$, we calculate the transport gaps for many anyon states between two Laughlin states $\nu=1/3$ and $\nu=1/5$. For $\nu_f=2$, we calculate the transport gaps for many anyon states between two Jain states $\nu=2/5$ and $\nu=2/9$.  As Fig.~\ref{gap} shows, the gaps vary smoothly with the change of $\theta$ and remain nonzero. This is a justification of the adiabatic heuristic principle proposed by Greiter and Wilczek \cite{Greiter90,Greiter92b}. A similar result is obtained by diagonalizing lattice Hamiltonian of smaller systems in Ref.~\cite{Kudo20}.

We note that we do not connect $\nu=1/3$ to $\nu=1$ or $\nu=2/5$ to $\nu=2$. The reason is technical: we are not able to perform the Jain-Kamilla projection for anyons in this filling factor region. However, in light of the above results, there is no reason to doubt that analogous adiabatic continuity in that filling factor range also holds.

\section{Chern numbers and Hall conductivity}
\label{Chern number}

In this section, we calculate the Chern number for the anyon wave function in Eq.~\ref{anyon wf} following the approach used by Niu, Thouless, and Wu~\cite{Niu85} and Tao and Haldane~\cite{Tao86}. The Chern number is defined as:
\be
\label{chern}
C=-i2\pi\sum_\alpha\langle{\partial J_2^{(\alpha)}\over \partial \phi_1}-{\partial J_1^{(\alpha)}\over \partial \phi_2}\rangle.
\ee
Here $\langle\rangle$ refers to the average in $(\phi_1,\phi_2)$ space, and the summation is over all degenerate ground states. $J_i^{(\alpha)}$ is defined as:
\be
J_i^{(\alpha)}=\sum_k\langle\Psi_k^{(\alpha)}|{\partial\over \partial \phi_i}|\Psi_k^{(\alpha)}\rangle
\ee
where the wave function is normalized, \ie $\sum_k\langle\Psi_k^{(\alpha)}|\Psi_k^{(\alpha)}\rangle=1$.
To see the periodicity of our wave function in the $(\phi_1,\phi_2)$ space, we need the identities Eq.~\ref{thetaa1} and Eq.~\ref{thetab1}.
Given these identities and the assumption that the overall normalization factor does not depend on $\phi_1$ and $\phi_2$, it is straightforward to see:
\begin{widetext}
\be
\label{C1}
\left(P_\alpha \Psi_{n;{p\over q}}\right)_k\left(\phi_1+2\pi(q+pn),\phi_2\right)=e^{i\pi (N-n)q}\left(P_\alpha \Psi_{n;{p\over q}}\right)_k\left(\phi_1,\phi_2\right)
\ee
\be
\label{C2}
\left(P_\alpha \Psi_{n;{p\over q}}\right)_k\left(\phi_1,\phi_2+2\pi(q+pn)\right)=e^{-i2\pi nq\left[{\phi_1\over 2\pi}-{p\over 2q}(N+1)+{pk\over q}+{N-n\over 2n}\right]}\left(P_\alpha \Psi_{n;{p\over q}}\right)_k\left(\phi_1,\phi_2\right)
\ee
Therefore, the average can be taken in the space $(0,2(q+pn)\pi)\otimes (0,2(q+pn)\pi)$. With the above identities, we can now prove
\ba
C&=&-{i2\pi\over(2\pi(q+pn))^2}\sum_\alpha\int_0^{2(q+pn)\pi}d\phi_1\int_0^{2(q+pn)\pi}d\phi_2\left[{\partial J_2^{(\alpha)}\over \partial\phi_1}-{\partial J_1^{(\alpha)}\over \partial \phi_2}\right]\\ \nonumber
&=&-{i2\pi\over(2\pi(q+pn))^2}\sum_\alpha\int_0^{2\pi (q+pn)}d\phi_2\left(J_2^{(\alpha)}(2(q+pn)\pi,\phi_2)-J_2^{(\alpha)}(0,\phi_2)\right)\\ \nonumber
&&+{i2\pi\over(2\pi(q+pn))^2}\sum_\alpha\int_0^{2\pi(q+pn)}d\phi_1\left(J_1^{(\alpha)}(\phi_1,2\pi (q+pn))-J_1^{(\alpha)}(\phi_1,0)\right)\\ \nonumber
&=&{qn}
\ea
\end{widetext}
The Chern number $qn$ depends not only on the fermionic filling factor $n$ but also on the statistical phase $\theta=\pi(1+p/q)$. Only when $\theta$ is an integer multiple of $\pi$ does the Chern number equal to $n$. Hence, it is not the Chern number but rather $C/q$ that remains invariant under the adiabatic evolution.  This is also consistent with the finding of Ref.~\cite{Kudo20}. (We note that the Chern number in Ref.~\cite{Kudo20} is actually equal to our $C/q$.) As shown in Ref.~\cite{Niu85}, the Hall conductivity in units of $e^2/h$ is the Chern number per degenerate ground state. Thereby, it is ${nq\over q+np}{e^2\over h}$ for our anyon states. 

\begin{figure}[t]
	\includegraphics[width=\columnwidth]{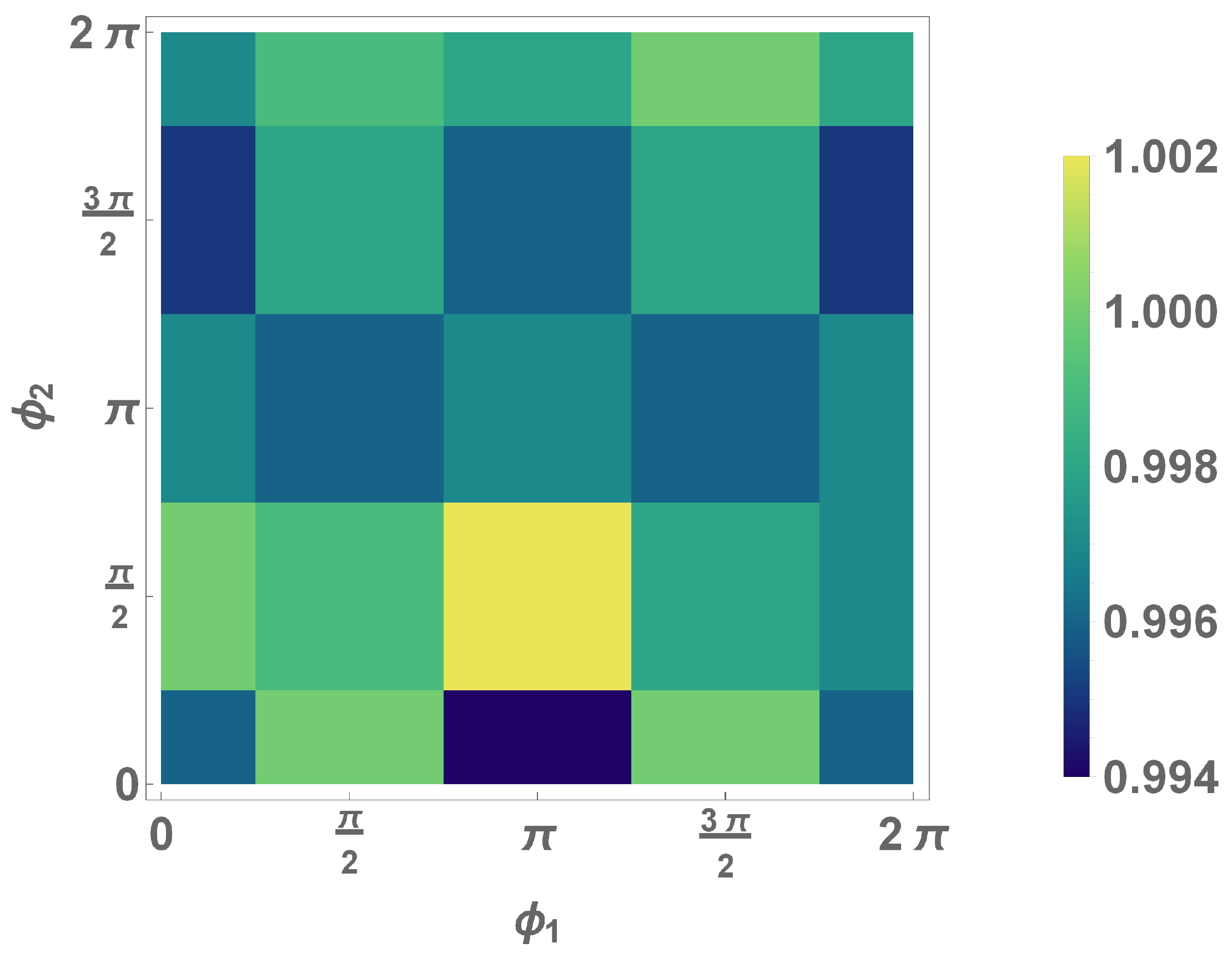} 
	\caption{
The Berry curvature divided by $q$, \ie $-i{2\pi\over q}\sum_\alpha \left({\partial J_2^{(\alpha)}\over \partial \phi_1}-{\partial J_1^{(\alpha)}\over \partial \phi_2}\right)$, at different points in the $(\phi_1,\phi_2)$ plane for our anyon wave function with $\nu_f=1$, $\nu=2/3$, $N=12$. The value is expected to be $1$.) We note the Berry curvature is uniform to a high degree, varying in a narrow range from $0.994$ to $1.002$. The sudden change in the color is an artifact, arising because the Berry curvature has been evaluated only for a discrete set of $(\phi_1,\phi_2)$ points.
 }
   \label{chern2D}
\end{figure}

\begin{figure}[t]
	\includegraphics[width=\columnwidth]{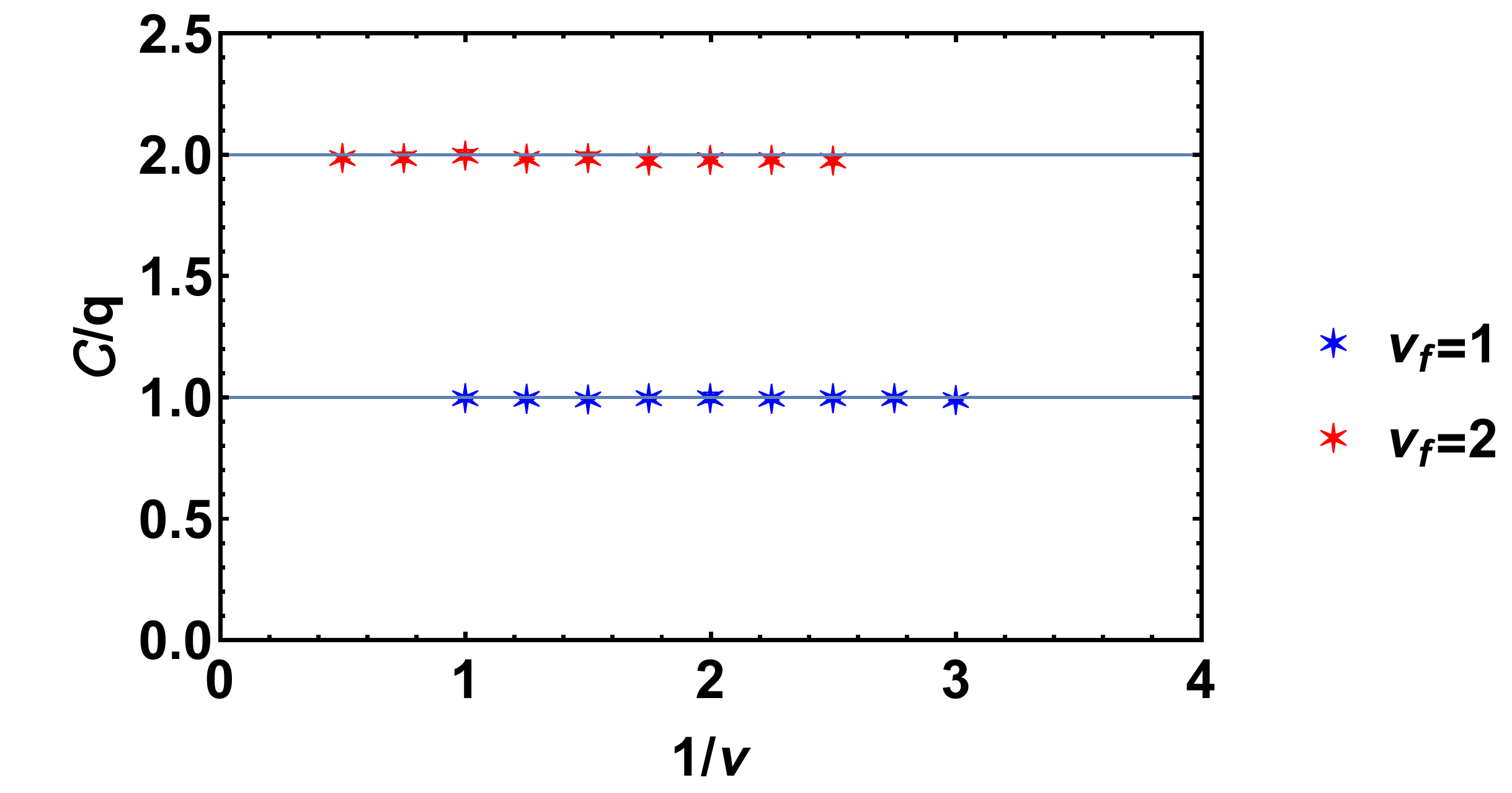} 
	\caption{
This figure shows $C/q=-i{2\pi\over q}\sum_\alpha \left({\partial J_2^{(\alpha)}\over \partial \phi_1}-{\partial J_1^{(\alpha)}\over \partial \phi_2}\right)$ evaluated at randomly chosen $(\phi_1,\phi_2)$ points for anyon ground states with $\nu_f=1,2$, $N=12$, where $C$ is the total Chern number. The values of $C/q$ are well quantized at $\nu_f$ as the statistics is varied, which agrees with the analytical result. The unprojected wave functions have been used for the calculation.
 }
   \label{chernno}
\end{figure}

As shown in Ref.~\cite{Kudo19} for fractional quantum Hall states, the integration or average over the twist angles in Eq.~\ref{chern} is not necessary when the system size is large enough, since the Berry curvature is already uniform. To see whether this is also true for anyon wave function, we calculate the Berry curvature at different points in the $(\phi_1,\phi_2)$ plane for $\nu_f=1$, $\nu=2/3$, $N=12$. As shown in Fig.~\ref{chern2D}, the Berry curvature is uniform to an extremely high degree, at the value derived above. We then calculate the $C/q=-i{2\pi\over q}\sum_\alpha \left({\partial J_2^{(\alpha)}\over \partial \phi_1}-{\partial J_1^{(\alpha)}\over \partial \phi_2}\right)$ without integration for different anyon wave functions with $\nu_f=1,2$, $N=12$. The results are shown in Fig.~\ref{chernno}. The numerical results are quantized at $\nu_f$, which agree with our analytical derivations above.

\section{Hall viscosities for anyons}
\label{Hall viscosities}
In addition to the Chern number, another topological quantity that can be easily calculated in torus geometry is the Hall viscosity. Avron, Seiler, and Zograf \cite{Avron95} showed that the Hall viscosity can be computed as Berry curvature through adiabatic deformation of the geometry of the torus: 
\be
\label{berry-curv}
\eta^A=-{\hbar \tau_2^2  \over V}\mathcal{F}_{\tau_1,\tau_2},
\ee
where 
\be
\label{BC}
\mathcal{F}_{\tau_1,\tau_2}=-2{\rm Im}\bigg\langle {\partial \Psi \over \partial \tau_1}\bigg|{\partial \Psi \over \partial \tau_2}\bigg\rangle.
\ee
Based on Eq.~\ref{berry-curv}, Read proposed \cite{Read09,Read10} that for fermionic and bosonic gapped states $\eta^A$ is given by
\be
\label{hall visc}
\eta^A=\sh{\hbar \rho \over 4}.
\ee
where $\rho=N/V$ and the ``shift" $\sh$ is a topological quantum number defined in the spherical geometry, given by $\sh={N\over \nu}-N_\phi$. 
This relation has been derived or numerically confirmed for Laughlin states, Pfaffian states, and Jain states by various approaches \cite{Read09,Read10,Tokatly09,Cho14,Fremling14,Lapa18,Lapa18b,Pu20}. In particular, Ref.~\cite{Pu20} developed an analytical derivation for microscopic wave functions. The main result of that work is that if a wave function is a product of several components, then the Hall viscosity is the sum of the Hall viscosities for different components provided that the normalization factor satisfies certain behavior in the thermodynamic limit. This statement holds for the unprojected as well as the projected Jain wave functions.

Clearly, Eq.~\ref{anyon wf2} is in a product form, and we can apply the theorem stated above. The fermionic part $\Psi_n$ contributes ${nN\hbar\over 4V}$ to the Hall viscosity. The remaining question is:  how much does the anyonic part $\Psi_{p\over q}$ contribute? As it is shown in Refs.~\cite{Tokatly09,Milovanovic10}, if the wave function (disregarding the normalization factor $\mathcal{N}$)  is a holomorphic function of $\tau$, which is the case for $\Psi_{p\over q}$, then its contribution to Hall viscosity is given by
\be
{\hbar \tau_2^2\over 2V}\left[\left({\partial\over \partial \tau_1}\right)^2+\left({\partial\over \partial \tau_2}\right)^2\right]\ln \mathcal{N}.
\ee
We further note that $\Psi_{p\over q}$ is very similar to the Laughlin wave function. They can both be separated into a center-of-mass part and a relative part [which is written in terms of $(z_i-z_j)$]. Furthermore, the relative part of $\Psi_{p\over q}$ has the same form as the relative part of the Laughlin wave function at $\nu=1/m$ with $m=p/q$. 
We now argue that the contribution of the center-of-mass to the Hall viscosity vanishes in the thermodynamic limit.  
The contribution of the center-of-mass part to $\ln \mathcal{N}$ is on the order of $\ln N$ ($N$ is the particle number), implying that its contribution to the Hall viscosity vanishes as $\ln N\over N$ in the thermodynamic limit. In fact, when deriving the Hall viscosity for Laughlin states, Tokatly and Vignale~\cite{Tokatly09} used the cylindrical geometry in which the center-of-mass part is absent, which also is valid only if the contribution of the center-of-mass part is unimportant. Hence, the total contribution of $\Psi_{p\over q}$ is ${p\rho\hbar\over 4q}$, and the Hall viscosity for the anyon wave function in Eq.~\ref{anyon wf2} is $\left(n+{p\over q}\right){\rho\hbar \over 4}$, or in terms of $\nu$ and $\nu_f$:
\be
\eta^A=\left({1\over \nu}-{1\over \nu_f}+\nu_f\right){\hbar \rho\over 4}.
\ee

We also calculate the Hall viscosity of the anyon wave function Eq.~\ref{anyon wf2} directly through Eq.~\ref{berry-curv}.  As mentioned above, the Hall viscosity is dominated by the relative part, so we only use the first component of the wave function and multiply the result by the number of components, just as we have done for energy. [We mention a slight subtlety in the calculation. Because of the presence of branch cuts in the Jastrow factors, we have to manually correct the jumps between different Riemann sheets. For instance, when we vary the geometry of the torus by a tiny amount, the imaginary part of $\ln \elliptic{1/2}{1/2}{{z_i-z_j\over L_1}}{\tau}$ might change by $\delta\pm2\pi$, where $\delta$ is a tiny number. In that case, we manually correct the change to $\delta$.]

\begin{figure}[t]
	\includegraphics[width=\columnwidth]{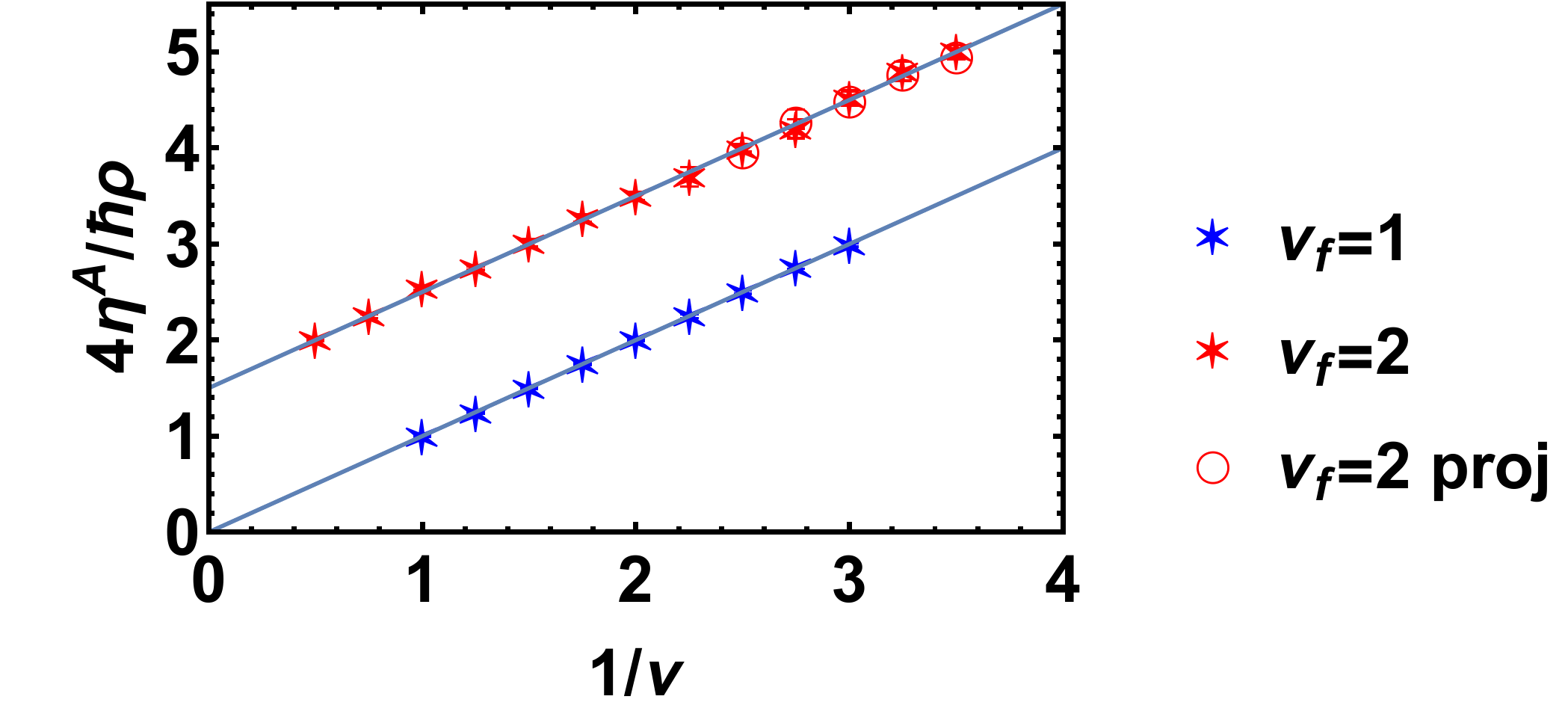} 
	\caption{
The Hall viscosity in units of $\hbar \rho\over 4$, for $\nu_f=1,2$. The straight lines indicate the theoretically predicted values, which are given by ${1\over \nu}$ and ${3\over 2}+{1\over \nu}$, respectively. The system contains 20 particles on a square torus. The $\nu_f=2$ results are for the unprojected wave functions (stars), but wherever possible, we have also evaluated the Hall viscosity for the LLL projected wave functions (circles). The $\nu_f=1$ wave functions are automatically in the LLL.
 }
	\label{hallvisc}
\end{figure}

The results are shown in Fig.~\ref{hallvisc}. We consider a system of 20 particles on a square torus $\tau=i$. We calculate the Hall viscosity at $\nu_f=1,2$ for different filling factors by varying  the statistical phase $\theta$. According to the analysis above, the Hall viscosity for $\nu_f=1$ is ${1\over \nu}{\rho\hbar \over 4}$ and for $\nu_f=2$ is $\left({3\over 2}+{1\over \nu}\right){\rho\hbar \over 4}$. The numerical results agree with these values. We also note that the unprojected wave functions and projected wave functions have the same Hall viscosity, as is also the case with the Jain states \cite{Pu20}.

\section{Summary}
\label{summary}
In summary, we have achieved a construction of multi-component anyon wave functions in the torus geometry. The wave functions are representations of the braiding group and have the expected ground state degeneracy. In the special cases in which $\theta$ is an integer multiples of $\pi$, the anyon wave functions return to the Jain CF wave functions \cite{Pu17}. When ${p\over q}\ge 2$, we can project the wave function to the LLL with the efficient modified Jain-Kamilla method. We calculate the transport gaps by evaluating the ground state, quasiparticle, and quasihole energies. The transport gap varies smoothly as we vary the statistical parameter $\theta$. We also calculate the Chern number, and we find that $C\over q$ is an adiabatic invariant, \ie it is invariant with the change of $\theta$. This is consistent with the exact diagonalization results of Kudo and Hatsugai\cite{Kudo20}. We also evaluate the Hall viscosity and find it to be $\left(n+{p\over q}\right){\rho\hbar \over 4}$ for $\nu_f=n,\theta=(1+{p\over q})\pi$. The results are summarized in Table.~\ref{table1}.

\begin{table*}
\makebox[\textwidth]{\begin{tabular}{|c|c|}
\hline
$\nu_f$&$n$\\ \hline
$\nu$&${nq\over np+q}$\\ \hline
$\theta$&$\pi\left(1+{p\over q}\right)$\\ \hline
number of components&$q$\\ \hline
ground state degeneracy&$q+np$\\ \hline
$C$&$nq$\\ \hline
Hall conductivity&${nq\over q+np}{e^2\over h}$\\ \hline
$\eta^A$&${\hbar \rho\over 4}\left({p\over q}+n\right)$\\ \hline
\end{tabular}}
\caption{\label{table1} Summary of the number of components, ground state degeneracy, Chern number ($C$), and Hall viscosity ($\eta^A$) of the anyon wave function at filling factor $\nu$ with statistical phase $\theta$ and effective filling $\nu_f$. The integer quantum Hall states correspond to $q=1$ and $p=0$, whereas the composite fermion states to $p=2s$ and $q=1$.}
\end{table*}

\begin{acknowledgments} 
We are grateful to Yayun Hu, Koji Kudo and Bin Wang for helpful discussions. This work was supported by the U. S. Department of Energy, Office of Basic Energy Sciences, under Grant No. DE-SC0005042. The numerical part of this research was conducted with Advanced CyberInfrastructure computational resources provided by the Institute for CyberScience at the Pennsylvania State University.
\end{acknowledgments}

\begin{appendix}

\section{Jacobi theta function with rational characteristics}
\label{theta-function}

The Jacobi theta function with rational characteristics\cite{Mumford07} is defined as
\be
\elliptic[\displaystyle]abz\tau=\sum_{n=-\infty}^{\infty}e^{i\pi \left(n+a\right)^2\tau}e^{i2\pi \left(n+a\right)\left(z+b\right)}.
\ee
The zeros of $\elliptic[\displaystyle]abz\tau$ lie at $z={1\over 2}-b+m+\left({1\over 2}-a+n\right)\tau$, where $m$ and $n$ are integers. We list here several periodic properties of Jacobi theta functions that have been used in our derivations in the main text:
\be
\elliptic[\displaystyle]ab{z+1}\tau=e^{i2\pi a}\elliptic[\displaystyle]abz\tau ,
\ee
\be
\elliptic[\displaystyle]ab{z+\tau}\tau=e^{-i\pi [\tau+2(z+b)]}\elliptic[\displaystyle]abz\tau ,
\ee
\be
\elliptic[\displaystyle]ab{z+w}{w\tau}=\elliptic[\displaystyle]a{b+w}z{w\tau} ,
\ee
\be
\label{branch}
\elliptic[\displaystyle]ab{z+\tau}{w\tau}=e^{-i{2\pi \over w}\left(z+b+{\tau\over 2}\right)}\elliptic[\displaystyle]{a+{1\over w}}bz{w\tau} ,
\ee
\be
\label{thetaa1}
\elliptic{a+1}{b}{z}{\tau}=\elliptic{a}{b}{z}{\tau},
\ee
\be
\label{thetab1}
\elliptic{a}{b+1}{z}{\tau}=e^{i2\pi a}\elliptic{a}{b}{z}{\tau},
\ee
where $w$ is a real number. 

\section{Distribution of the twisted boundary phases}
\label{appen a}
In this appendix, we discuss the constrain on the distribution of the twisted boundary phases $\phi_i^a$ and $\phi_i^f$ ($i=1,2$) for composite anyon wave functions. As mentioned in the main text and Ref.~\cite{Pu17}, the preservation of periodic boundary conditions only require $\phi_i=\phi_i^a+\phi_i^f$. This implies, in general: 
\be
\phi_i^a=\gamma\phi_i,\quad i=1,2,
\ee
\be
\phi_i^f=(1-\gamma)\phi_i,\quad i=1,2,
\ee
where $\gamma$ is a real number. We fix $\gamma$ as follows. 
The phases correspond to the effective magnetic field through the holes of the torus felt by the attached vortices and fermions, which are in the proportion $np/q$. Requiring the same proportion for the phases yields
$\gamma={np\over np+q}$. In this context, we note that the 
wave function in Eq.~\ref{C1} and Eq.~\ref{C2} produces a different Chern number for $\gamma\neq {np\over np+q}$; 
the reason is that then the wave function $P_\alpha\Psi_{n;{p\over q}}$ vanishes for some values of $\phi_1$ and $\phi_2$. Hence, the calculation of Chern only makes sense for $\gamma={np\over np+q}$. We next demonstrate this statement 
for $n=1$, leaving 
the generalization to other values of $n$ as an open question.

The anyon wave function Eq.~\ref{anyon wf2} can be written as a superposition of different momentum eigenstates.
\be
\label{decom}
\Psi_{1;{p\over q}}=\Psi_1\Psi_{p\over q}^{(N-1)}=\sum_{\alpha=0}^{p+q-1}c_\alpha \Psi_{1+{p\over q}}^{(\alpha)}
\ee
Here $\Psi_{1+{p\over q}}^{(\alpha)}$ are the momentum eigenstates. They can be obtained by applying $P_\alpha$ defined in Eq.~\ref{P}. For the special case $n=1$, there is an easier way: they are simply given by Eq.~\ref{anyon wf} with the replacement $p'\rightarrow p+q$. Our task is to show all $c_\alpha$ are nonzero for arbitrary $\phi_1$ and $\phi_2$ only when $\gamma={p\over p+q}$.

The Jastrow factors on the left-hand-side and right-hand-side of Eq.~\ref{decom} are obviously identical. The coefficients are 
hence determined by the center-of-mass part. In other words, Eq.~\ref{decom} can be rewritten as:
\begin{widetext}
\be
\label{decom2}
\elliptic{a^f}{b^f}{Z\over L_1}{\tau}\elliptic{a^a_{k}}{b^a}{Z\over L_1}{q\tau\over p}\sim\sum_{\alpha=0}^{p+q-1} c_\alpha \elliptic{a_k}{b_\alpha}{Z\over L_1}{q\tau\over p+q}
\ee
\end{widetext}
where the parameters are given by:
\be
a^f={N-1\over 2}+{\phi_1^f\over 2\pi}
\ee
\be
b^f={N-1\over 2}-{\phi_2^f\over 2\pi}
\ee
\be
a^a_k={1\over 2\pi}\left(\phi_1^a-{p\pi\over q}(N+1)+2\pi {pk\over q}\right)
\ee
\be
b^a=-{1\over 2\pi}\left({q\over p}\phi_2^a+\pi(N+1)\right)
\ee
\be
a_k={1\over 2\pi}\left(\phi_1-{p+q\over q}\pi(N+1)+2\pi{p+q\over q}k\right)
\ee
\be
b_\alpha=-{1\over 2\pi}\left({q\over p+q}\left(\phi_2+2\pi \alpha\right)-2\pi (N-1){q\over p}+\pi (N+1)\right)
\ee
In Eq.~\ref{decom2} we use $\sim$ instead of $=$ since, for simplicity, we 
have omitted normalization factors in this equation and below. 
This does not influence our judgment whether $c_\alpha$ is zero. 
To solve for $c_\alpha$, we expand the theta function according to its definition and compare the coefficients of $e^{i2\pi(n+a_3){Z\over L_1}}$ on left-hand-side and right-hand-side:
\begin{widetext}
\be
c_\alpha e^{i\pi(n+a_k)^2{q\tau\over p+q}+i2\pi(n+a_k)b_\alpha}\sim \sum_{m_1,m_2}\delta_{n+a_k,m_1+m_2+a^f+a^a_k}e^{i\pi(m_1+a^f)^2\tau+i\pi(m_2+a^a_k)^2{q\tau\over p}+i2\pi(m_1+a^f)b^f+i2\pi(m_2+a^a_k)b^a}.
\ee
\end{widetext}
Since $c_\alpha$ cannot depend on $n$ or $k$, we 
choose $n=0,k=0$. With some algebra, we get
\be
c_\alpha\sim \elliptic{a^f-a_0{q\over p+q}}{b^f-b^a}{0}{{q+p\over p}\tau}.
\ee
The condition for $c_\alpha=0$ is
\be
a^f-a_0{q\over p+q}={\phi_1\over 2\pi}\left({p\over p+q}-\gamma\right)+N={1\over 2}+l_1
\ee
\be
b^f-b^a={\phi_2\over 2\pi}\left({q+p\over p}\gamma-1\right)+N={1\over 2}+l_2
\ee
Here $l_1$ and $l_2$ are two arbitrary integers and $N$ is the particle number. 
When $\gamma={p\over p+q}$ 
it is not possible to satisfy the above two equations for any $\phi_1$ and $\phi_2$. 
For $\gamma\neq {p\over p+q}$, one can always find values of $\phi_1$ and $\phi_2$ to make $c_\alpha$ zero. This proves our statement that we have to choose $\gamma={p\over p+q}$ to ensure that the wave function remains non-zero in the entire $(\phi_1,\phi_2)$ space.

\section{Modular covariance of the anyon wave functions}
\label{appen b}
As mentioned in the main text, the geometry of a torus is parameterized by $L_1$ and $L_2$. However, the parametrization is not unique. The geometry is unchanged under a modular transformation of $L_1$ and $L_2$, 
\be
\bigl(
\begin{smallmatrix}
L_2'\\
L_1'
\end{smallmatrix}
\bigr)
=
\bigl(
\begin{smallmatrix}
a &b\\
c &d
\end{smallmatrix}
\bigr)
\bigl(
\begin{smallmatrix}
L_2\\
L_1
\end{smallmatrix}
\bigr),
\ee
where $a,b,c,d\in\mathbb{Z}$ with $ad-bc=1$. These transformations form the modular group, which is spanned by two elements $\mathcal{T}$: $\bigl(\begin{smallmatrix}
1 &1\\
0 &1
\end{smallmatrix}\bigr)$ and $\mathcal{S}$: $\bigl(\begin{smallmatrix}
0 &-1\\
1 &0
\end{smallmatrix}\bigr)$. The geometry itself is unchanged by these transformations. If we redefine the twisted periodic boundary phases $\phi_1$ and $\phi_2$ consistently with the modular transformations, all the physical quantities should be invariant under modular transformations. To guarantee this, the wave functions must be covariant under modular transformations. To be more specific, the Hilbert space of the degenerate ground states is invariant under modular transformations, and the transformations of degenerate ground states are described by a unitary matrix. In this appendix, we show that our anyon wave functions do possess theses properties.

First let us consider the case in which there is no degeneracy, \ie $p=1$. In this case, the ground state is nondegenerate and thus should be invariant under modular transformation. This is true for fermions and bosons, whose wave functions are single-component. However, for anyons, the transformation is more subtle. Under a $\mathcal{S}$ transformation, the new $L_1$ direction is the original $L_2$ direction. Thereby, $t_i(L_1)$ acting on the original wave function is now represented by a nondiagonal matrix instead of a diagonal matrix. One can, however, recover the forms of Eq.~\ref{bg1} and Eq.~\ref{bg2} by performing a unitary transformation on the original ground state wave function $\Psi_k$:
\be
\tilde{\Psi}_k=\sum_{k'}U_{kk'}\Psi_{k'}\quad k,k'=0,1,2\cdots q-1
\ee
In other words, the components are mixed and reordered. The matrix can be 
obtained by comparing the periodic properties of $\tilde{\Psi}_k$ and $\Psi_k$. The matrices for $\mathcal{T}$ and $\mathcal{S}$ are
\be
U_{kk'}\left(\mathcal{T}\right)={1\over \sqrt{q}}\delta_{kk'}c^{k(k-1)\over 2}
\ee
\be
U_{kk'}\left(\mathcal{S}\right)={1\over \sqrt{q}}c^{-kk'}
\ee
with $c=e^{i2\pi{p'\over q}}$ as defined in the main text.

When the ground state degeneracy is present, there is another set of matrices  $V$ that describes the mixing of degenerate ground states $\Psi^{(\alpha)}=P_\alpha \Psi_{n;{p\over q}}$ under modular transformation. The $V$ matrices can be derived 
by comparing the properties of $\Psi^{(\alpha)}$ under $t_{\rm CM}\left(L_1\over N_\phi\right)$ and $t_{\rm CM}\left(L_2\over N_\phi\right)$ before and after the modular transformations. The matrices for $\mathcal{T}$ and $\mathcal{S}$ are
\be
V_{\alpha \alpha'}\left(\mathcal{T}\right)={1\over\sqrt{np+q}}e^{i\left({2\pi(\alpha-\alpha')\over np+q}+\theta_0\right){\alpha'-\alpha\over nq}}
\ee
\be
V_{\alpha \alpha'}\left(\mathcal{S}\right)={1\over\sqrt{np+q}}e^{i\left({\theta_1\alpha-\theta_2\alpha'\over nq}-{2\pi\alpha\alpha'\over nq(np+q)}\right)}
\ee
where $\theta_0={\pi q(nN_\phi-N)\over np+q}$, $\theta_1={\pi q(N-n)-n\pi (N+1)(p+q)\over np+q}$, and $\theta_2={\pi q(N-n)+n\pi (N+1)(p+q)\over np+q}$. The final modular transformation is described by the direct products of $U$ and $V$:
\be
\tilde{\varphi}_k^{(\alpha)}=U_{kk'}V_{\alpha \alpha'}\varphi_{k'}^{(\alpha')}
\ee
Since the direct product $U\otimes V$ is unitary, the physical quantities are guaranteed to be invariant under modular transformations.
As shown in Ref.~\cite{Fremling19}, the modified Jain-Kamilla projection preserves the modular covariance of the wave functions.

\begin{figure}[t]
	\includegraphics[width=\columnwidth]{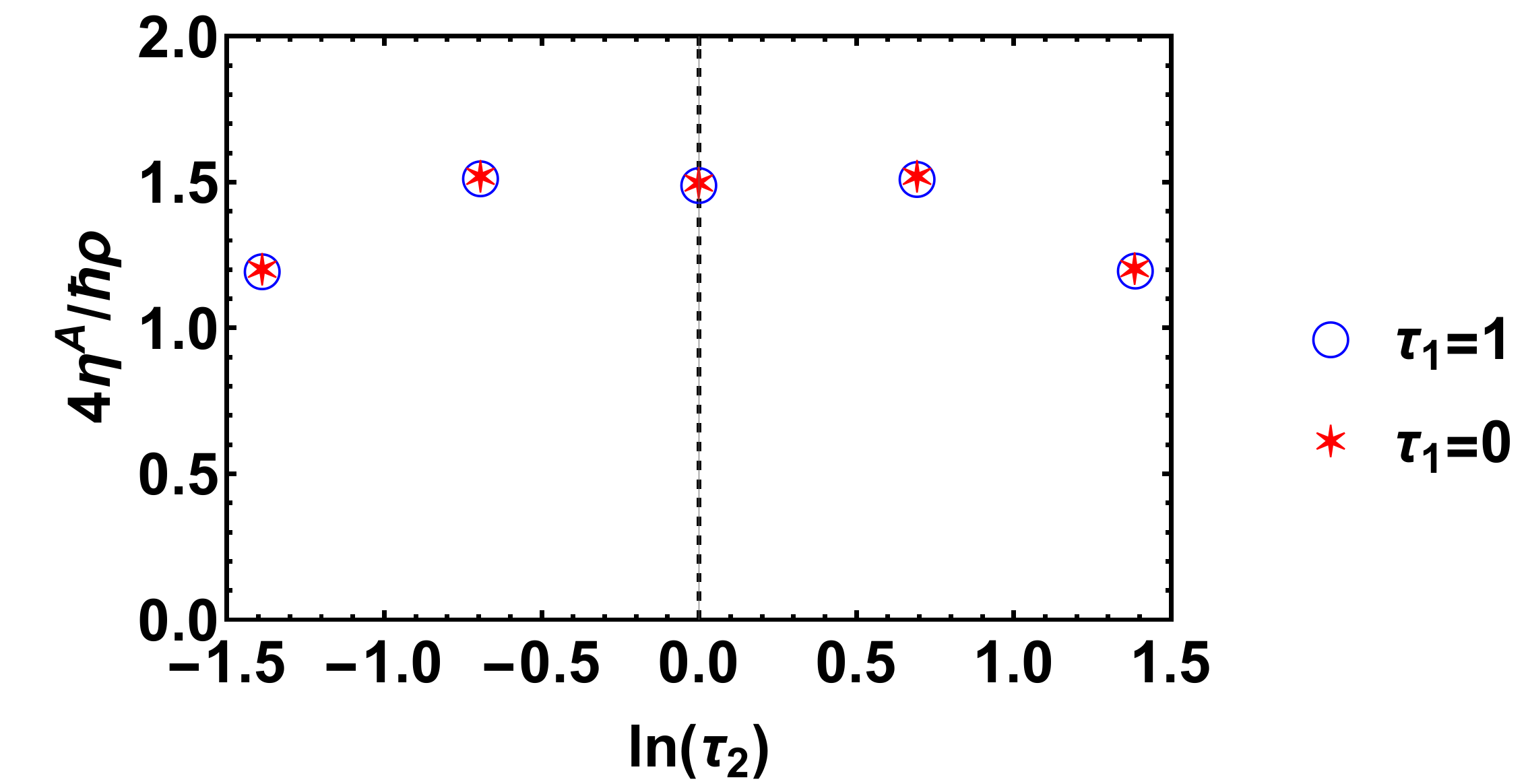} 
	\caption{
The Hall viscosity for different shapes of torus in units of $\hbar \rho\over 4$, for $\nu_f=1,\theta=3\pi/2,N=4$. The red and blue symbols are related by the $T$ transformation, and the left-hand side and right-hand side of the $y$-axis are related by the $S$ transformation. 
 }
	\label{modcov}
\end{figure}

To confirm the modular covariance of the wave functions numerically,  we calculate the Hall viscosities for wave functions that are related by modular transformations. The result is shown in Fig.~\ref{modcov}. We choose $\nu_f=1,\theta=3\pi/2,N=4$. As a result of the covariance under $T$ transformation, blue circles and red stars are supposed to be coincident, and the data are expected to be symmetric with respect to the $y$-axis because of the covariance under the $S$ transformation. 
The numerical results in Fig.~\ref{modcov} 
are explicitly consistent with these expectations, 
thus demonstrating that the wave functions are modular covariant.

\end{appendix}

\bibliographystyle{apsrev}

\end{document}